\documentclass{ws-ijmpa}

\begin{document}

\markboth{Gonz\'alez, Valcarce, Vijande, Garcilazo}{$b \bar b$ description
with a screened potential}

\catchline{}{}{}{}{}

\title{$b\bar b$ DESCRIPTION WITH A SCREENED POTENTIAL}

\author{P. Gonz\'alez}
\address{Dpto. de F\'\i sica Te\'orica and IFIC, Universidad de
Valencia - CSIC \\ E-46100 Burjassot, Valencia, Spain}

\author{A. Valcarce, J. Vijande}
\address{Grupo de F\'\i sica Nuclear, Universidad de Salamanca \\
E-37008 Salamanca, Spain}  

\author{H. Garcilazo}
\address{Escuela Superior de F\'\i sica y Matem\'aticas, Instituto
Polit\'ecnico Nacional, Edificio 9 \\ 07738 M\'exico D.F., Mexico}  

\maketitle

\pub{Received (Day Month Year)}{Revised (Day Month Year)}


\vspace*{0.4cm}
There has been a significant progress in the last decade in deriving
precisely, from lattice calculations, the static interaction potential
between two heavy quarks. In the so-called quenched approximation (only
valence, no sea quarks to start with) a funnel potential containing a
linear plus a color-coulomb term is well established~\cite{Bali}. Spin and
velocity dependent corrections to this form have been obtained as 
well~\cite{BSW}. When sea quarks are incorporated (unquenched approximation) 
the long-distance behavior may change dramatically. As a matter of fact it has
been shown in QCD at finite temperature~\cite{DeTar} and also in SU(2)
Yang-Mills theory \cite{FP} that the potential saturates, i.e.,
it gets a constant value from a certain distance. Physically the
saturation of the potential is related to screening: light $q\overline{q}$
pairs are created out of the vacuum between the two heavy-quark sources
giving rise to a screening of their color charges. The effect coming out at
short distances is that the running of the QCD coupling slows down with the
distance. This translates in an effective coulomb strength bigger than in
the quenched case. At long distances (bigger than the saturation distance)
string breaking may take place: it
becomes energetically favorable the recombination of the light $q$ and 
$\overline{q}$ of the pair with the two heavy sources to form two
static-light mesons.

Although saturation has not been proven in QCD~\cite{Bali} one can 
wonder whether there
is any direct indication of it from real QCD and/or
phenomenology. We shall argue that this question has a positive answer. Then
we shall construct a potential model incorporating screening for the
description of hadrons. Its application to the non-relativistic bottomonium
system will allow us to study the phenomenological effect of screening and
to predict a lower bound for the saturation distance.

From real QCD the Schwinger-Dyson equation allows to extract the running
coupling, $\alpha _{s}$, behavior in terms of an effective gluon mass,
$M_{g}(Q^{2})$~\cite{AG}. This behavior is connected to confinement.
Actually when this coupling, particularized for $M_{g}(Q^{2}\rightarrow
0)\sim \Lambda _{QCD}$, is implemented into the one-gluon-exchange diagram a
confining linear potential comes out. More generally a certain connection
can be unveiled by assuming a Yukawa parametrization of the confining
potential as proposed long time ago from lattice simulations~\cite{BL}:
\begin{equation}
\overline{V}_{conf}(r)=\overline{\sigma }r\left( \frac{1-e^{-\mu r}}{\mu r}%
\right) =\frac{\overline{\sigma }}{\mu }-\overline{\sigma }r\frac{e^{-\mu r}%
}{\mu r}  \, .
\end{equation}

By realizing that the linear behavior is recovered for $\mu $, the inverse
of the saturation distance, going to $0,$ the following connection has been
proposed~\cite{GVGV}:
\begin{equation}
\mu (Q^{2})=\Lambda _{QCD}-M_{g}(Q^{2})  \, .
\end{equation}
Then plausible values for the saturation distance, $\mu ^{-1},$ for heavy
meson scales, go from
$\mu ^{-1}(Q^{2}=(5$ GeV)$^{2})=1.25$ fm to $\mu ^{-1}(Q^{2}=(1$ GeV)$^{2})
=2.15$ fm.

From phenomenology the finite number of experimentally detected hadronic
resonances could well be the reflection of the saturation property of the
potential. Indeed from a calculation of the non-strange baryon spectra with
a Yukawa-screened interaction a perfect one-to-one correspondence between
quark model bound states and experimental resonances can be 
obtained~\cite{VGGV}. For heavy mesons the same parametrization
allows for an adequate description of $b\overline{b}$ and to a lesser extent
of $c\overline{c}.$ However this smooth parametrization does not seem to
correspond to more recent lattice indications of a rather abrupt saturation
transition~\cite{FAS}. To check the 
effect of such a fast deconfinement we shall
analyze its consequences in the simplest non-relativistic
system, bottomonium, by using the potential model:
\begin{equation}
V(r)=\left\{\matrix{ \sigma r-e/r & & r < r_{br} \cr
\sigma r_{br}-e/r_{br} & & r \geq r_{br}}\right. \, ,
\end{equation}
where $\sigma$ is a confining parameter whose value can be guessed from the
value of the force at intermediate distances~\cite{SO} $(\sigma \sim
800-900$ MeV$\cdot$fm$^{-1})$, $e$ is the unquenched Coulomb strength
$(e=106$ MeV$\cdot$fm $(\alpha_s=0.4))$,
and $r_{br}$ is the breaking or saturation distance for which we expect 
1 fm $< r_{br}< $ 2 fm at the $(Q^{2})_{b\overline{b}}$ scale. In order to be
more precise we fit the $b\overline{b}$ spectrum. By choosing the $b$ quark
mass altogether with $\sigma$ so as to reproduce the ground state mass and
spectral energy separations $(m_b=4820.5$ MeV, $\sigma =800$ 
MeV$\cdot$fm$^{-1})$, $r_{br}$ is fitted by assuming that the
energy breaking or saturation threshold, $E_{br}$, is just above the highest
energy resonance known, say $E_{br}\sim 11050$ MeV. In this manner a lower
bound value for $r_{br}$ $(r_{br}=1.76$ fm) is in fact
obtained. The results for the complete spectrum (up to $d$ states) are shown
in Table 1 where the calculated masses are assigned to spin-triplet states.
Since differences in masses with data are at most of 30 MeV an unambiguous
identification of states is possible though for the $ns$ states with 
$n\geq 4$ some room for a slight mixing with $d$ states is left. 

\begin{table}[h]
\tbl{$b \bar b$ spin-triplet bound state masses and properties.}
{\begin{tabular}{@{}ccccc@{}} \toprule
& Mass (MeV) & Exp. & $\left\langle {v^{2}/c^{2}}\right\rangle $ & $%
\left\langle r^{2}\right\rangle ^{1/2}$ (fm) \\ \colrule
$1s$ & 9460  & $9460.30\pm 0.26$ & 0.10 & 0.21 \\ 
$2s$ & 10022 & $10023.26\pm 0.31$ & 0.08 & 0.51 \\ 
$1d$ & 10171 & $10162.2\pm 1.6$ & 0.02 & 0.56 \\ 
$3s$ & 10346 & $10355.2\pm 0.5$ & 0.09 & 0.76 \\ 
$2d$ & 10444 &  & 0.04 & 0.80 \\ 
$4s$ & 10605 & $10580.0\pm 3.5$ & 0.10 & 0.97 \\ 
$3d$ & 10679 &  & 0.06 & 1.00 \\ 
$5s$ & 10829 & $10865\pm 8$ & 0.11 & 1.12 \\ 
$4d$ & 10889 &  & 0.07 & 1.23 \\ 
$6s$ & 11030 & $11019\pm 8^{\,}$ & 0.11 & 1.43 \\ \colrule
$1p$ & 9935 & $9900.1\pm 0.5$ & 0.03 & 0.41 \\ 
$2p$ & 10267 & $10260.0\pm 0.5$ & 0.05 & 0.67 \\ 
$3p$ & 10531 &  & 0.07 & 0.89 \\ 
$4p$ & 10761 &  & 0.08 & 1.09 \\ 
$5p$ & 10958 &  & 0.08 & 1.43\\ \botrule
\end{tabular}}
\end{table}

This is confirmed
by the calculation of leptonic widths, Table 2 (for details of the
calculation see Ref.~\refcite{GVGV}), showing 
a good agreement with experiment (see
the fourth column where the effect of radiative and relativistic corrections
is partially eliminated for $ns$ states by taking their ratio to the $1s$
case). 

\begin{table}[h]
\tbl{Leptonic widths $\Gamma_{e^+e^-}$ for $b\bar b$ in keV.}
{\begin{tabular}{@{}ccccc@{}} \toprule
& $\Gamma _{e^{+}e^{-}}^{(0)}\left( 1-{\frac{{16\alpha _{s}}}{{3\pi }}}%
\right) $ & $\left( \Gamma _{e^{+}e^{-}}\right) _{\rm exp}$ & $\Gamma
_{e^{+}e^{-}}^{(0)}/\Gamma _{e^{+}e^{-}}^{(0)}(1s)$ & $\left[ \Gamma
_{e^{+}e^{-}}/\Gamma _{e^{+}e^{-}}(1s)\right] _{\rm exp}$ \\ \colrule
$1s$ & 1.01 & $1.32\pm 0.05$ & 1 & 1 \\ 
$2s$ & 0.35 & $0.520\pm 0.032$ & 0.35 & $0.41\pm 0.07$ \\ 
$1d$ & $2.2\times 10^{-4}$ &  & $2.2\times 10^{-4}$ &  \\ 
$3s$ & 0.25 & seen & 0.25 & seen \\ 
$2d$ & $2.6\times 10^{-4}$ &  & $2.6\times 10^{-4}$ &  \\ 
$4s$ & 0.22 & $0.248\pm 0.031$ & 0.22 & $0.19\pm 0.03$ \\ 
$3d$ & $5.7\times 10^{-4}$ &  & $5.7\times 10^{-4}$ &  \\ 
$5s$ & 0.18 & $0.31\pm 0.07$ & 0.18 & $0.24\pm 0.06$ \\ 
$4d$ & $6.7\times 10^{-4}$ &  & $6.7\times 10^{-4}$ &  \\ 
$6s$ & 0.14 & $0.130\pm 0.030$ & 0.14 & $0.10\pm 0.03$\\ \botrule
\end{tabular}}
\end{table}

Electromagnetic E1 transitions have been also evaluated, Table 3, in
excellent agreement with data. 

\begin{table}[h]
\tbl{E1 decay widths for $b \bar b$ in keV.}
{\begin{tabular}{@{}ccccc@{}} \toprule
Transition & $\Gamma _{E1}$ & $\Gamma _{\rm exp}$ &  &  \\ \colrule
$\Upsilon (2s)\rightarrow \gamma \chi _{b_{0}}(1P)$ & 1.62 & $1.7\pm 0.5$ & 
&  \\ 
$\Upsilon (2s)\rightarrow \gamma \chi _{b_{1}}(1P)$ & 2.55 & $3.0\pm 0.8$ & 
&  \\ 
$\Upsilon (2s)\rightarrow \gamma \chi _{b_{2}}(1P)$ & 2.51 & $3.1\pm 0.8$ & 
&  \\ \colrule
$\Upsilon (3s)\rightarrow \gamma \chi _{b_{0}}(2P)$ & 1.77 & $1.4\pm 0.4$ & 
&  \\ 
$\Upsilon (3s)\rightarrow \gamma \chi _{b_{1}}(2P)$ & 2.88 & $3.0\pm 0.6$ & 
&  \\ 
$\Upsilon (3s)\rightarrow \gamma \chi _{b_{2}}(2P)$ & 3.14 & $3.0\pm 0.6$ & 
&\\ \botrule 
\end{tabular}}
\end{table}

Finally the spin singlet-triplet splitting
has been calculated perturbatively. Our result $M\left[ \Upsilon (1s)\right]
-M\left[ \eta _{b}(1s)\right] \simeq 185.7$ MeV agrees with the
experimental value\ $M\left[ \Upsilon (1s)\right] -M\left[ \eta _{b}(1s)%
\right] =160\pm 40$ MeV within the experimental errors.

We can then conclude that our naive non-relativistic quark model provides 
with an excellent description of bottomonium, much better than the
corresponding to a non-screened funnel potential and at the level of the
best descriptions obtained with much more ellaborated models. These results
seem to support the view that a rather abrupt saturation may take place in
real QCD. Moreover we predict that at the momentum scale corresponding to 
$b\overline{b}$ the saturation distance should be bigger than 1.75 fm.
From this large value one does not expect saturation can be easily seen in
current lattice simulations with the usual Wilson loops techniques.
Meanwhile it could be worthwhile to examine at the phenomenological level
the effects of an abrupt saturation in other meson and baryon systems.


\section*{Acknowledgements}
This work was partially funded by Direcci\'{o}n General de
Investigaci\'{o}n Cient\'{\i}fica y T\'{e}cnica (DGICYT) under the Contract
No. BFM2001-3563, by Junta de Castilla y Le\'{o}n under the Contract No.
SA-104/04, by EC-RTN, Network ESOP, contract HPRN-CT-2000-00130, by
Oficina de Ciencia y Tecnolog\'\i a de la Comunidad Valenciana, Grupos03/094,
and by COFAA-IPN (Mexico).



\begin{thebibliography}{0}
\bibitem{Bali} G. S. Bali, 
	{\it Phys. Rep.} {\bf 343}, 1 (2001).

\bibitem{BSW} G. S. Bali, K. Schilling, and A. Wachter,
	{\it Phys. Rev.} {\bf D56}, 2566 (1997).

\bibitem{DeTar} C. DeTar, O. Kaczmarek, F. Karsch, and E. Laermann, 
	{\it Phys. Rev.} {\bf D59}, 031501 (1999).

\bibitem{FP} P. de Forcrand and O. Philipsen, 
	{\it Phys. Lett.} {\bf B475}, 280 (2000).

\bibitem{AG} A. C. Aguilar, A. Mihara, and A. A. Natale, 
	{\it Phys. Rev.} {\bf D65}, 054011 (2002).

\bibitem{BL} K. D. Born, E. Laermann, N. Pirch, T. F. Walsh, and P. M. Zerwas, 
	{\it Phys. Rev.} {\bf D40}, 1653 (1980).

\bibitem{GVGV} P. Gonz\'alez, A. Valcarce, H. Garcilazo, and J. Vijande,
	{\it Phys. Rev.} {\bf D68}, 034007 (2003).

\bibitem{VGGV} J. Vijande, P. Gonz\'alez, H. Garcilazo, and A. Valcarce,
	{\it Phys. Rev.} {\bf D69}, 074019 (2004).

\bibitem{FAS} G. S. Bali {\it et al.}, 
	{\it Phys. Rev.} {\bf D62}, 2566 (1997).

\bibitem{SO} R. Sommer, 
	{\it Nucl. Phys.} {\bf B411}, 839 (1994).
\end{thebibliography}
\end{document}